\documentstyle[12pt]{article}
\newcommand{\dd}{\mbox{\rm d}}
\newcommand{\nnn}{\noindent}
\newcommand{\p}{\partial}
\newcommand{\be}{\begin{equation}}
\newcommand{\ee}{\end{equation}}
\newcommand{\bi}{\bibitem}

\begin{document}
\thispagestyle{empty}
\title{THE EMBEDDING MODEL OF INDUCED GRAVITY\\[5pt]
       WITH BOSONIC SOURCES}

\author{Matej Pav\v si\v c\
\\
       Jo\v zef Stefan Institute, University of Ljubljana, Ljubljana,
       Slovenia}
\maketitle

\abstract{
We consider a theory in which spacetime is an n-dimensional surface $V_n$
embedded in an $N$-dimensional space $V_N$. In order to enable also
the Kaluza-Klein approach we admit $n > 4$. The dynamics is given by
the minimal surface action in a curved embedding space. The latter is taken,
in our specific model, as being a conformally flat space. In the quantization
of the model we start  from a generalization of the Howe-Tucker action
which depends on the embedding variables ${\eta}^a (x)$ and the
(intrinsic) induced metric $g_{\mu \nu}$ on $V_n$. If in the path integral
we perform only the functional integration over ${\eta}^a (x)$, we obtain
the effective action which functionally depends on $g_{\mu \nu}$ and
contains the Ricci scalar $R$ and its higher orders $R^2$ etc. But due
to our special choice of the conformal factor in $V_N$ enterig our original
action, it turns out that the effective action contains also the source term.
The latter is in general that of a $p$-dimensional membrane ($p$-brane);
in particular we consider the case of a point particle. Thus, starting from
the basic fields ${\eta}^a (x)$, we induce not only the kinetic term for
$g_{\mu \nu}$, but also the "matter" source term.}

\medskip

email: {MATEJ.PAVSIC@IJS.SI}

\newpage

\section{Introduction}

After so many years of intensive research quantization of gravity is
still un unfinished project. Among many approaches followed there is the
one which seems to be especially promising. This is the so called induced
gravity
proposed by Sakharov \cite{1}. His idea was to treat metrics
not as a fundamental field but as one induced from more basic fields. The idea
has been pursued by numerous authors \cite{2}; especially illuminating
are works by Akama, Terasawa and Naka \cite{3}. Their basic action contains $N$
scalar fields and it is formally just is a slight generalization of the well
known Dirac-Nambu-Goto action for an $n$-dimensional worldsheet swept
by an $(n-1)$-dimensional membrane.

In the present paper we are going to pursue such an approach and
give a concrete physical interpretation to the $N$-scalar fields which we
shall denote ${\eta}^a (x)$. We shall
assume that spacetime is a surface $V_4$, called spacetime sheet,
embedded in a higher dimensional space $V_N$, and ${\eta}^a (x)$ are
the embedding functions. An embedding model has been
first proposed by Regge and Teitelboim \cite{4} and investigated by
others \cite{5}. In that model the action contains the Ricci scalar
expressed in terms
of the embedding functions. In our present model \cite{6}, \cite{6a}
on the contrary, we
start from the action which is essentially the minimal surface action weighted
with a function $\omega (\eta)$ in $V_N$. For a suitably chosen $\omega$,
such that it is singular ($\delta$-function like) on certain surfaces
${\hat V}_m$, also embedded in $V_N$, we obtain on $V_4$ a set of worldlines.
It is shown that these worldlines are geodesics of $V_4$ provided that
$V_4$, described by ${\eta}^a (x)$, is a solution to our variational procedure.
I shall show that after performing functional integrations over
${\eta}^a (x)$ we obtain two contributions to the path integral. One
contribution comes from all possible ${\eta}^a (x)$ not intersecting
${\hat V}_m$ and the other contribution from those ${\eta}^a (x)$ which
do intersect the surfaces ${\hat V}_m$. In the effective action so obtained,
the first contribution gives the Einstein-Hilbert term $R$ plus higher order
terms like $R^2$. The second contribution can be cast into the form of
the path integral over all possible worldlines $X^{\mu} (\tau)$. We so
obtain the action which contains matter sources and the kinetic term for
the metric field (plus higher orders in $R$). So in the proposed approach
both, the metric field and the matter field  are induced from more basic
fields ${\eta}^a (x)$.

\section{The classical model}

We assume that the arena in which physics takes place is an $N$-dimensional
space $V_N$ with $N \geq 10$. Next we assume that an $n$-dimensional
surface $V_n$ living in $V_n$ represents a possible spacetime. The parametric
equation of such a "spacetime sheet" $V_n$ is given by the embedding
functions ${\eta}^a (x^{\mu})$, $a = 0,1,2,...,N$, where $x^{\mu}$ ,
$\mu = 0,1,2,...,n-1$ are coordinates (parameters) on $V_n$. We suppose
that the action is just the one for a minimal surface $V_n$
\be
   I[{\eta}^a (x)] = \int (\mbox{det} \, {\p}^{\mu} {\eta}^a \, {\p}^{\nu}
  {\eta}^b
  {\gamma}_{ab})^{1/2} {\dd}^n x
\label{1}
\ee

\nnn where ${\gamma}_{ab}$ is the metric tensor of $V_N$. Dimension of a
spacetime sheet $V_n$ is taken here to be arbitrary, in order to allow
for the Kaluza-Klein approach. In particular we may take $n = 4$. We admit
that the embedding space is curved in general. In particular let us consider
 the case of a conformally flat $V_N$ such that ${\gamma}_{ab} =
{\omega}^{2/n} {\eta}_{ab}$, where ${\eta}_{ab}$ is the $N$-dimensional
Minkowski tensor. Then Eq.(\ref{1}) becomes
\be
   I[{\eta}^a (x)] = \int  \omega (\eta )
   (\mbox{det} \, {\p}^{\mu} {\eta}^a \, {\p}^{\nu} {\eta}^b
  {\eta}_{ab})^{1/2} {\dd}^n x
\label{2}
\ee

\nnn From now on we shall forget about the origin of $\omega (\eta )$ and
we shall consider it as a function of position in a {\it flat} embedding space.
Indices $a,b,c$ will be raised and lowered by ${\eta}^{ab}$ and ${\eta}_{ab}$,
respectively.

In principle $\omega (\eta )$ is arbitrary. But it is very instructive to
choose
the following function
\be
    \omega (\eta ) = {\omega}_0 + \sum_i \int m_i {{{\delta}^N (\eta -
   {\hat \eta}_i)}
    \over {\sqrt{|\gamma|}}} \, \, {\dd}^m {\hat x} \sqrt{|{\hat f}|}
\label{3}
\ee

\nnn where ${\eta}^a = {\hat \eta}_i^a ({\hat x})$ is the parametric equation
of
an $m$-dimensional surface ${\hat V}_m^{(i)}$, called {\it matter sheet}, also
embedded in $V_N$, ${\hat f}$ is the determinant of the induced metric
on ${\hat V}_m^{(i)}$, and $\sqrt{|\gamma|}$ allows for taking curved
coordinates in otherwise flat $V_N$. If we take $m = N - n + 1$, then the
intersection of $V_n$ and ${\hat V}_m^{(i)}$ can be a (one-dimensional)
line, i.e. a worldline $C_i$ on $V_n$. In general, when $m = N - n + (p+1)$,
the intersection can be a $(p+1)$-dimensional worldsheet representing motion
of a $p$-dimensional membrane (also called $p$-brane). In this paper we
confine our consideration to the case $p = 0$, that is to motion of a point
particle.

Inserting (\ref{3}) into (\ref{2}) and writing $f_{\mu \nu} \equiv
{\p}^{\mu} {\eta}^a {\p}^{\nu} {\eta}_a \; , \; f \equiv \mbox{det} \,
f_{\mu \nu}$
we obtain
\be
   I[\eta] = {\omega}_0 \, \int {\dd}^n x \, \sqrt{|f|} + \int {\dd}^n x
  \sum_i m_i \,
  \delta^n (x - X_i) (f_{\mu \nu} {\dot X}_i^{\mu} {\dot X}_i^{\nu})^{1/2}
  \, {\dd}
  \tau
\label{4}
\ee

\nnn The above result was obtained by writing ${\dd}^m {\hat x} = {\dd}^{n-1}
{\hat x} \, {\dd}^{n-1} \tau$, ${\hat f} = {\hat f}^{(m-1)} ({\dot X}_i^{\mu}
{\dot X}_{i \mu})^{1/2}$ and taking the coordinates ${\eta}^a$
such that ${\eta}^a = (x^{\mu}, {\eta}^n, ... , {\eta}^{N-1})$,
where $x^{\mu}$ are (curved) coordinates on $V_n$. The determinant of the
metric of the
embedding space $V_N$ in such curvilinear coordinates is then
$\gamma = \mbox{det} \, {\p}^{\mu} {\eta}^a \, {\p}^{\mu} {\eta}_a = f$.

If we vary the action (\ref{4}) with respect to ${\eta}^a (x)$ we obtain
\be
   {\p}_{\mu} \left [ \sqrt{|f|} \, ({\omega}_0 f^{\mu \nu} + T^{\mu \nu})
   {\p}_{\nu} {\eta}_a \right ] = 0
\label{5}
\ee

\nnn where
\be
   T^{\mu \nu} = {1 \over {\sqrt{|f|}}} \, \sum_i \int {\dd}^n x \, m_i \,
   {\delta}^n (x - {\dot X}_i ) \, {{{\dot X}_i^{\mu} {\dot X}_i^{\nu}} \over
   {({\dot X}_i^{\alpha} {\dot X}_{i \alpha} )^{1/2}}} \, {\dd} \tau
\label{6}
\ee

\nnn is the stress-energy tensor of dust. Eq.(\ref{5}) can  be rewritten in
terms
of the covariant derivative $D_{\mu}$ on $V_n$:
\be
    D_{\mu} \left [ ({\omega}_0 f^{\mu \nu} + T^{\mu \nu})
   {\p}_{\nu} {\eta}_a \right ] = 0
\label{7}
\ee

\nnn The latter equation gives
\be
   {\p}_{\nu} {\eta}_a \, D_{\mu} T^{\mu \nu} + ({\omega}_0 f^{\mu \nu} +
  T^{\mu \nu}) D_{\mu} D_{\nu} {\eta}_a = 0
\label{8}
\ee

\nnn where we have taken into account that covariant derivative of metric is
zero,
i.e. $D_{\alpha} f_{\mu \nu} = 0$, and $D_{\alpha} f^{\mu \nu} = 0$, which
implies also ${\p}_{\alpha} \eta^c \, D_{\mu} D_{\nu} {\eta}_c = 0$, since
$f_{\mu \nu} \equiv {\p}_{\mu} {\eta}^a {\p}_{\nu} {\eta}_a $. Contracting
Eq.(\ref{8}) by ${\p}^{\alpha} {\eta}^a $ we have
\be
    D_{\mu} T^{\mu \nu} = 0
\label{9}
\ee

\nnn The latter are the well known equations of motion for sources. In the
case of dust (\ref{9}) implies that dust particles move along geodesics of
spacetime $V_n$. We have thus obtained a very interesting result that the
worldlines $C_i$ which are obtained as the intersections $V_n \cap
{\hat V}_m^{(i)}$ are geodesics of the spacetime sheet $V_n$. The same
result is obtained also directly by varying the action (\ref{4}) with respect
to variables ${\dot X}_i^{\mu} (\tau)$.

A solution to the equations of motion (\ref{5}) (or (\ref{7})) gives both: a
spacetime sheet ${\eta}^a (x)$ and worldlines ${\dot X}_i^{\mu} (\tau)$.
Once ${\eta}^a (x)$ is determined, also the induced metric
$g_{\mu \nu} = {\p}_{\mu} {\eta}^a {\p}_{\nu} {\eta}_a$ is determined. But
such a metric, in general, does not satisfy Einstein's equations. In the next
section we shall see that quantum effects induce the necessary Einstein-
Hilbert term $(-g)^{1/2} R$.

\section{The quantum model}

For the purpose of quantization we shall use a classical action \cite{6a}
which ia a generalization of the well known Howe-Tucker action
\cite{Howe-Tucker}
which is equivalent to (\ref{2}):
\be
   I[{\eta}^a , g^{\mu \nu}] = {1 \over 2} \int {\dd}^n x \sqrt{|g|} \, \omega
  (\eta ) (g^{\mu \nu} {\p}_{\mu} {\eta}^a {\p}_{\nu} {\eta}_a + 2 - n)
\label{10}
\ee

\nnn It is a functional of the embedding functions ${\eta}^a (x)$ and the
Lagrange multipliers $g^{\mu \nu}$. Varying (\ref{10}) with respect to
$g^{\mu \nu}$ gives the constraints
\be
   - {{\omega} \over 4} \, \sqrt{|g|} \, g_{\alpha \beta}
  (g^{\mu \nu} {\p}_{\mu} {\eta}^a {\p}_{\nu} {\eta}_a + 2 - n) + {{\omega}
  \over 2} \, \sqrt{|g|} \, {\p}_{\alpha} {\eta}^a {\p}_{\beta} {\eta}_a = 0
\label{11}
\ee

\nnn Contracting (\ref{11}) with $g^{\alpha \beta}$ we find
$g^{\mu \nu} {\p}_{\mu} {\eta}^a {\p}_{\nu} {\eta}_a = n$, and after inserting
the latter relation back into (\ref{11}) we find
\be
   g_{\alpha \beta} = {\p}_{\alpha} {\eta}^a {\p}_{\beta} {\eta}_a
\label{12}
\ee

\nnn which is the expression for the induced metric on a surface $V_n$. In the
following paragraphs we shall specify $n = 4$; however, whenever necessary
we shall switch to a generic case of arbitrary $n$.

In the classical theory we may say that a 4-dimensional spacetime sheet is
swept by a 3-dimensional space-like hypersurface $\Sigma$ which moves
forward in time. The latter surface is specified by initial conditions and
equations of motion then determine $\Sigma$ at every value of a time-like
coordinate $x^0 = t$. Knowledge of a particular hypersurface $\Sigma$
implies knowledge of the corresponding intrinsic 3-geometry given by the
3-metric $g_{ij} = {\p}_i {\eta}^a {\p}_j {\eta}_a$ induced on $\Sigma$
$(i, j = 1,2,3)$. However, knowledge of data ${\eta}^a (t, x^i)$ on an entire
infinite $\Sigma$ is just a mathematical idealization which cannot be realized
in a practical situation by an observer, because of the finite speed of light.

In quantum theory a state of a surface $\Sigma$ is not given by coordinates
${\eta}^a (t, x^i)$, but by a wave functional $\psi [t, {\eta}^a (x^i)]$. The
latter represents probability amplitude that at time $t$ an observer
would get, as a result of measurement, a particular surface $\Sigma$.

The probability amplitude for the transition from a state with definite
${\Sigma}_1$
at time $t_1$ to a state ${\Sigma}_2$ at time $t_2$ is given by the Feynman
path integral
\be
   K(2,1) = \langle {\Sigma}_2 , t_2 | {\Sigma}_1 , t_1 \rangle =
   \int e^{i \, I[\eta , g]}
   {\cal D} \eta \, {\cal D} g
\label{13}
\ee

\nnn Now, if in Eq.(\ref{13}) we perform integration only over the embedding
functions ${\eta}^a (x^{\mu})$, then we obtain the so called {\it effective
action} $I_{eff}$
\be
    e^{i I_{eff} [g]} \equiv \int e^{i \, I[\eta , g]}  {\cal D} \eta
\label{14}
\ee

\nnn which is functional solely of the metric $g^{\mu \nu}$. From Eq.(\ref{14})
we
obtain by functional differentiation
\be
  {{\delta I_{eff} [g]} \over {\delta g^{\mu \nu}}} = {{\int  {{\delta I_
[\eta,
   g]} \over {\delta g^{\mu \nu}}} \, e^{i \, I[\eta , g]}  {\cal D} \eta }
   \over {\int e^{i \, I[\eta , g]}  {\cal D} \eta }}
\equiv \langle {{\delta I [\eta, g]} \over {\delta g^{\mu \nu}}} \rangle= 0
\label{15}
\ee

\nnn On the left hand side of Eq.(\ref{15}) we have taken into account the
constraints
${{\delta I [\eta, g]} \over {\delta g^{\mu \nu}}} = 0$ (explicitly given in
Eq.(\ref{11}).

The expression ${{\delta I_{eff} [\eta, g]} \over {\delta g^{\mu \nu}}} = 0$
gives the
the classical equations for the metric $g_{\mu \nu}$, derived from the
effective
action.

Let us now consider a specific case in which we take for $\omega (\eta) $ the
expression (\ref{3}). Then our action (\ref{10}) splits into two terms
\be
I[\eta,g] = I_0 [\eta, g] + I_m [\eta , g]
\label{16}
\ee

\nnn  with
\be
 I_0 [\eta , g] = {{{\omega}_0} \over 2} \int {\dd}^n \, x \sqrt{|f|} \,
        (g^{\mu \nu} {\p}_{\mu} {\eta}^a {\p}_{\nu} {\eta}_a + 2 - n)
\label{17}
\ee
\be
  I_m [\eta , g] = {1 \over 2} \, \int {\dd}^n x \, \sqrt{|g|} \,
  \sum_i \, m_i \,
  {{{\delta}^N (\eta - {\hat \eta}_i )} \over {\sqrt{|\gamma|}}} \,
  {\dd}^m {\hat x}
  \, \sqrt{|{\hat f}|} \, (g^{\mu \nu} {\p}_{\mu} {\eta}^a {\p}_{\nu}
  {\eta}_a + 2 - n)
\label{18}
\ee

\nnn The last expression can be integrated over $m - 1$ coordinates
${\hat x}^{\mu}$, while ${\hat x}^0$ is chosen so to coincide with a parameter
$\tau$ of a worldline $C_i$. We also split the metric as $g^{\mu \nu} =
{{n^{\mu} n^{\nu}}/{n^2}} + {\bar g}^{\mu \nu}$, where $n^{\mu}$ is
a time-like vector and ${\bar g}^{\mu \nu}$ the projection tensor, giving
${\bar g}^{\mu \nu} {\p}_{\mu} {\eta}^a {\p}_{\nu} {\eta}_a = n - 1$.
So we obtain
\be
   I_m [\eta , g] = {1 \over 2} \int {\dd}^n x \, \sqrt{|g|} \, \sum_i \,
  {{{\delta}^n (x - X_i (\tau ))} \over
  { \sqrt{|g|}}} \, \left ( {{g_{\mu \nu} {\dot X}_i^{\mu} {\dot X}_i^{\nu}}
 \over {{\mu}_i}}
 + {\mu}_i \right ) = I_m [{\dot X}_i , g]
\label{19}
\ee

\nnn Here ${\mu}_i \equiv 1/ \sqrt{n^2} |_{C_i}$ are the Lagrange multipliers
giving, after variation, the worldline constraints ${\mu}_i^2 = {\dot X}_i^2$.
Equation (\ref{19}) is the well known Howe-Tucker action \cite{Howe-Tucker} for
point particles.

Now let us substitute our specific action (\ref{16})-(\ref{19}) into the
expression
(\ref{14}) for the effective action. The functional integration now runs over
two distinct classes of spacetime sheets $V_n$ [represented by
${\eta}^a (x)$]:

(a) those $V_n$ which either {\it do not intersect} the matter sheets
${\hat V}_m^{(i)}$ [represented by ${\hat \eta}_i^a ({\hat x})$], or if they
do,
the intersections are just single points, and

(b) those $V_n$ which {\it do intersect} ${\hat V}_m^{(i)}$, the intersections
being worldlines $C_i$.

\nnn The sheets $V_n$ which correspond to the case (b) have two
distinct classes of points (events):

(b1) {\it the points outside the intersection}, i.e., outside the worldlines
$C_i$ ,

(b2) {\it the points on the intersection}, i.e., the events belonging to $C_i$.

\nnn The measure ${\cal D} \eta^a (x)$ can be factorized into the contribution
which corresponds to the case (a) or (b1) ($x \notin C_i$), and into the
contribution which corresponds to the case (b2) ($x \in C_i$):
 $$  {\cal D} \eta = \prod_{a,x} (|g(x)|)^{1/4} {\dd} {\eta}^a (x)$$
\be
  \quad \quad = \prod_{a,x \notin C_i} (|g(x)|)^{1/4} {\dd} {\eta}^a (x)
   \prod_{a,x \in C_i} (|g(x)|)^{1/4} {\dd} {\eta}^a (x) \equiv
  {\cal D}_0 \eta \, {\cal D}_m  \eta
\label{20}
\ee

\nnn
The additional factor $(|g(x)|)^{1/4}$ comes from the requirement that the
measure be
invariant under reparametrizations of $x^{\mu}$ (see Ref. \cite{mera} for
details).
{}From the very definition of $\prod_{a,x \in C_i} (|g(x)|)^{1/4} {\dd}
{\eta}^a (x)$
as the measure of the set of points on the worldlines $C_i$ [each $C_i$ being
represented by an equation $x = {\dot X}_i^{\mu} (\tau)$] we conclude that
\be
  {\cal D}_m {\eta}^a (x) = {\cal D} {\dot X}_i^{\mu} (\tau)
\label{21}
\ee

\nnn The effective action then satisfies [using (\ref{16})-(\ref{21})]:
\be
   e^{i I_{eff} [g]} = \int e^{i I_0 [\eta , g]} \, {\cal D}_0 \eta \, \,
  e^{i I_m [{\dot X}_i , g ]} \, {\cal D} {\dot X}_i \equiv e^{i W_0}\,
  e^{i W_m}
\label{22}
\ee
\be
   I_{eff} = W_0 + W_m
\label{23}
\ee

\nnn The measure ${\cal D}_0 \eta$ includes all those sheets $V_n$ that
do not intersect a matter sheet [case (a)], and also all those which do
intersect [case (b1)], apart from the points on ${\hat V}_m^{(i)}$
[case (a)].

The first factor in the product (\ref{22}) contains the action (\ref{10}). The
latter has the same form as the action for $N$ scalar fields in a curved
background spacetime with the metric $g_{\mu \nu}$. The corresponding
effective action has been studied and derived Refs. \cite{Birrel}. Using the
same procedure and taking our specific constants ${\omega}_0/2$ and
$(n - 1)$ occurring in Eq.(\ref{10}) we find for the effective Lagrangian
the following expression:
\be
   L_{eff} = n \ {\omega}_0^{-1} (4 \pi )^{- n/2} \sum_{j=0}^{\infty}
   (n - 2)^{n/2 - j} \, a_j (x) \, \Gamma (j - {n \over 2})
\label{24}
\ee

\nnn  with
\begin{eqnarray}
  a_0 (x) & = & 1 \\
  a_1 (x) & = & {R \over 6} \\
  a_2 (x) & = & {1 \over 12} R^2 + {1 \over 180} (R_{\alpha \beta \gamma
  \delta} R^{\alpha \beta \gamma  \delta} - R_{\alpha \beta} R^{\alpha
  \beta}) - {1 \over 20} D_{\mu} D^{\mu} R
\end{eqnarray}

\nnn where $R, R_{\alpha \beta}$ and $R_{\alpha \beta \gamma \delta} $
are the Ricci scalar, the Ricci tensor and the Riemann tensor, respectively.
The function $\Gamma (y) = \int_0^{\infty} e^{-t} \, t^{y - 1} \, {\dd} t$;
it is divergent at negative integers $y$ and finite at $y = {3 \over 2},
{1 \over 2}, - {1 \over 2}, - {3 \over 2}, - {5 \over 2}, ...$ .The effective
Langrangian (\ref{24}) is thus divergent in even dimensional spaces $V_n$.
For instance, when $n = 4$, the argument in Eq.(\ref{24}) is $j - 2$ which,
for $j = 0, 1, 2, ...$ , is indeed a negative integer. Therefore, in order to
obtain
a finite effective action in 4-dimensions one needs to introduce a suitable
cut off parameter, so that  $L_{eff}$ depends on that parameter. On the
contrary, in an odd dimensional space $L_{eff}$ is finite and has the form
\be
   L_{eff} = {\lambda}_0 + {\lambda}_1 R + {\lambda}_2 R^2 + {\lambda}_3
  (R_{\alpha \beta \gamma
  \delta} R^{\alpha \beta \gamma  \delta} - R_{\alpha \beta} R^{\alpha
  \beta}) + {\lambda}_4 D_{\mu} D^{\mu} R
\label{28}
\ee

\nnn where
$$   {\lambda}_0 = K \, \Gamma(- {n \over 2}) \quad , \quad \;
   {\lambda}_1 = {K \over {6 (n - 2)}} \, \Gamma (1 - {n \over 2}) \quad ,
\quad \; {\lambda}_2 = {K \over {12 (n - 2)^2}} \, \Gamma (2 - {n \over
  2} )$$
\be
{\lambda}_3 = {K \over {180 (n - 2)^3}} \, \Gamma (3 - {n \over 2}) \quad ,
\quad \quad {\lambda}_4 = - \, {K \over {20 (n - 2)^4}} \, \Gamma (4- {n \over
  2} )
\label{29}
\ee
\nnn and $K \equiv {\omega }_0^{-1} n (n - 2)^{n/2} (4 \pi )^{- n/2}$.
\nnn For instance, when $n = 5$, we have $K = {\omega}_0^{-1} 3 \,
{\left ( {3 \over {4 \pi}} \right )}^{5/2}$, and
\be
{\lambda}_0 = - {{\sqrt{3}}
\over {{\pi}^2 {\omega}_0}} \; , \; {\lambda}_1 = - {5 \over {36} } \,
{\lambda}_0 \; , \;
{\lambda}_2 = {5 \over {144}} \, {\lambda}_0 \; , \; {\lambda}_3 = {5 \over
{2160}} \, {\lambda}_0 \; , \; {\lambda}_4 = - {5 \over {240}} \, {\lambda}_0
\label{30}
\ee

\nnn Here ${\lambda}_0 $ is the cosmological constant, while ${\lambda}_1$
is related to the gravitational constant $G$ in $n$ dimensions according
to ${\lambda}_1 \equiv (16 \pi G)^{-1}$. In 5 dimensions we have from
(\ref{30}) that $(16 \pi G)^{-1} = {{\sqrt{3}} \over {{\pi}^2 {\omega}_0}}
{5 \over {36}}$. This last relation shows how the induced (5-dimensional)
gravitational constant is calculated in terms of ${\omega}_0$ which is
a free parameter of our embedding model.

So, if take seriously the Kaluza-Klein theories in which spacetime has more
than 4 dimensions, then it makes sense to consider a spacetime sheet
$V_n$ of an odd dimension $n$ = 5, 7 or 9, etc., which leads straightforwardly
to a finite effective action, without need to introduce a cut off parameter.
Such
a higher dimensional effective action can then be reduced to 4 dimensions
by taking the extra dimensions compactified on a very small length
(e.g. Plank length).

In the above calculation of the effective action we have considered all
functions ${\eta}^a (x)$ entering the path integral (\ref{14}) as representing
distinct spacetimes sheets $V_n$. However, because of the
reparametrization invariance there exist equivalence classes of functions
representing the same $V_n$. This complication must be taken into account,
and the conventional approach is to introduce ghost fields which cancel
the unphysical degrees of freedom. An alternative approach, explored in
Ref.\cite{7} is to assume that all possible embedding functions ${\eta}^a (x)$
can be nevertheless interpreted to describe physically distinct
spacetime sheets ${\cal V}_n$. This is possible if we the extra degrees
of freedom in ${\eta}^a (x)$ describe deformations of ${\cal V}_n$. Such
a deformable surface ${\cal V}_n$ is then a different concept than a
non-deformable surface $V_n$. The path integral can be straightforwardly
performed in the case of ${\cal V}_n$, as we did it in arriving at the
result (\ref{24}).

Let us now return to Eq.(\ref{22}). In the second factor of Eq.(\ref{22})
the functional integration runs over all possible worldlines $X^{\mu} (\tau)$.
Though they are obtained as intersections of {\it various} $V_n$ with
${\hat V}_m^{(i)}$, we may consider all those worldlines as lying in
{\it the same} effective spacetime $V_n^{(eff)}$ with the intrinsic metric
$g_{\mu \nu}$. In other words, in the effective theory, we identify all those
various $V_n$'s, having the same induced (intrinsic) metric $g_{\mu \nu}$,
to be one and the same spacetime. If one considers the embedding space
$V_N$ of sufficiently high dimension $N$, then there is enough freedom
to obtain as the intersection any possible worldline in the effective
spacetime $V_n^{(eff)}$.

When the condition for the classical approximation is stisfied, i.e., when
$I_m >> \hbar = 1$, then only those trajectories ${\dot X}_i^{\mu}$ which
are close to the classically allowed ones effectively contribute:
\be
   e^{i W_m} = e^{i I_m [X_i, g]} \, \quad , \quad \quad W_m = I_m
\label{31}
\ee

The effective action is then the sum of the gravitational field kinetic term
$W_0$ given in Eq.(\ref{28}) and the source term $I_m$ given in (\ref{19}).
Variation of $I_{eff}$ with respect to $g^{\mu \nu}$ then gives the
equations of the gravitational field in the presence of point-particle sources
with the stress energy tensor $T^{\mu \nu}$ as given in Eq(\ref{6}):
\be
    R^{\mu\nu} - {1 \over 2} g^{\mu \nu} + {\lambda}_0 g^{\mu \nu} +
   (\mbox{\rm higher order terms}) = - 8 \pi G \, T^{\mu \nu}
\label{32}
\ee

However, in general the classical approximation is not satisfied, and in the
eva\-luation of the matter part $W_m$ of the effective action one must take
into account the contribution of all possible paths ${\dot X}_i^{\mu} (\tau)$.
So we have (confining to the case of only one particle, omitting the
subscript $i$, and taking $\mu = 1$)
\be
e^{i W_m} = \int_{x_0}^{x_b} {\cal D} X \, \mbox{exp} \left ( {i \over 2}
 \int_{{\tau}_a}^{{\tau}_b} {\dd} \tau \, m \, (g_{\mu \nu} {\dot X}_i^{\mu}
{\dot X}_i^{\nu} + 1) \right ) = {\cal K} (x_b , {\tau}_b ; x_a , {\tau}_a)
\equiv {\cal K} (b,a)
\label{33}
\ee

\nnn which is the propagator or the Green function satisfying (for
${\tau}_b \ge {\tau}_a$)
\be
  \left ( i\, {{\p} \over {\p {\tau}_b}} - H \right )
  {\cal K} (x_b , {\tau}_b ; x_a , {\tau}_a) = - {1 \over {\sqrt{|g|}}} \,
  {\delta}^n
  (x_b - x_a ) \, \delta ({\tau}_b - {\tau}_a )
\label{34}
\ee

\nnn where $H = (|g|)^{-1/2} {\p}_{\mu} ((|g|)^{1/2} {\p}^{\mu} )$. From
(\ref{34})
we have
\be
    {\cal K} (b,a) = - i \, {\left [ i {{\p} \over {\p {\tau}_b}} -
     H \right ]}_
{x_b , {\tau}_b ; x_a , {\tau}_a}^{-1}
\label{35}
\ee

\nnn where the inverse Green function is treated as a matrix in $(x, \tau)$
space.

Using the following relation \cite{Kaku} for Gaussian integration
\be
    \int y_m y_n \prod_{i = 1}^N {\dd} y_i {\dd} y_j \, e^{- y_i A_{ij} y_j}
  \propto {{(A^{-1})_{mn}} \over {(\mbox{\rm det} \, |A_{ij}|^{1/2}}}
\label{36}
\ee
\nnn we can rewrite the Green function in terms of the second quantized
field
\be
  {\cal K} (a,b) = \int {\psi }^* (x_b, {\tau}_b ) \, {\psi } (x_b, {\tau}_b )
  \,
  {\cal D} {\psi}^* \, {\cal D} \psi \, \, \mbox{\rm exp} \left [ -i
   \int {\dd} \tau \,
  {\dd}^n x \, \sqrt{|g|} \, {\psi}^* (i {\p}_{\tau} - H) \psi \right ]
\label{37}
\ee

If the conditions for "classical" approximation are satisfied, such that
the phase in (\ref{37}) is much greater than $\hbar = 1$, then only those
paths $\psi (\tau, x) \; , \quad {\psi}^* (\tau , x )$ which are close to the
extremal path, along which the phase is zero, effectively contribute to
${\cal K} (a , b)$. Then the propagator is simply
\be
  {\cal K} (b , a) \propto \mbox{\rm exp} \, \left [  -i \, \int {\dd} \tau \,
  {\dd}^n x \, \sqrt{|g|} \, {\psi}^* (i {\p}_{\tau} - H) \psi \right ]
\label{38}
\ee

\nnn The effective, one-particle, "matter" action $W_m$ is then
\be
   W_m = -  \int {\dd} \tau \,
  {\dd}^n x \, \sqrt{|g|} \, {\psi}^* (\tau , x) (i {\p}_{\tau} - H) \psi
  (\tau , x)
\label{39}
\ee

\nnn If we assume that $\tau$-dependence of the field $\psi (\tau , x) $ is
given by $\psi (\tau , x) = e^{i m \tau} \, \phi (x)$, then Eq.(\ref{39})
simplifies
to the usual well known expression for a scalar field:
 $$W_m = \int {\dd}^n x \, \sqrt{|g|} \, {\phi}^* (x) \left ( {1
 \over {\sqrt{|g|}}}
  \, {\p}_{\mu} (\sqrt{|g|} \, {\p}^{\mu} ) + m^2 \right ) \phi (x)$$
\be
  \quad \quad = - {1 \over 2} \int {\dd}^n x \, \sqrt{|g|} \, (g^{\mu \nu} \,
  {\p}_{\mu} {\phi}^* \, {\p}_{\nu} \phi - m^2 )
\label{40}
\ee

\nnn where the surface term has been omitted.

Starting from our basic fields ${\eta}^a (x)$ which are the embedding
functions for a spacetime sheet $V_n$ we arrived at the effective action
$I_{eff}$ which contains the kinetic term $W_0$ for the metric field $g^{
\mu \nu}$ (see Eq.(\ref{28})) and the source term $W_m$ (see Eq.(\ref{32})
and (\ref{18}), or (\ref{40})). Both, the metric field $g_{\mu \nu}$ and
the bosonic matter field $\phi $ are induced from the basic fields
${\eta}^a (x)$.

\section{Conclusion}

We have investigated a model which seems to be very promising in attempts
to find a consistent relation between quantum theory and gravity. Our
model exploits the approach of induced gravity and the concept of
embedding spacetime in a higher dimensional space and has the following
interesting property: what appears as worldlines in e.g. 4-dimensional
spacetime
are just the intersections of a spacetime sheet $V_4$ with "matter"
sheets  ${\hat V}_m^{(i)}$. Various choices of spacetime sheets then give
various configurations od worldlines. Instead of $V_4$ it is convenient to
consider a spacetime sheet $V_n$ of arbitrary dimension $n$. When passing
to the quantized
theory, a spacetime sheet is no longer definite. All possible alternative
spacetime sheets are taken into account in an expression for a wave
functional or a Feynman path integral. The points of intersection of a $V_n$
with a matter sheet ${\hat V}_m^{(i)}$ are treated specially, and it is found
that their contribution to a path integral is identical to the contribution of
a point-particle path. We have paid special attention to the effective action
which results after having functionally integrated out all possible embeddings
which give the same induced metric tensor. We have found that the
effective action, besides the Einstein-Hilbert term and corresponding
higher-order terms, contains also the source term. The expression for
the latter is equal to that of a classical (when such an approximation can
be used) or quantum point-particle source described by a scalar (bosonic)
field.

In other words, we have found that ($n$-dimensional) Einstein's equations
(including $R^2$
and higher derivative terms) with classical or quantum point-particle
sources are effective equations resulting after having performed the
quantum average over all possible embeddings of spacetime. Gravity - as
described by Einstein's general relativity - is thus considered not as
fundamental phenomenon, but as being induced quantum mechanically from
more fundamental phenomena.

In our embedding model of gravity with bosonic sources, new and
interesting possibilities are open. For instance, instead of a 4-dimensional
spacetime sheet we can consider a sheet which possesses additional
dimensions, parametrized either with the usual or the Grassmann
coordinates. In such a way we expect to include, on the one hand, via the
Kaluza-Klein mechanism, also other interactions besides the gravitational
one, and on the other hand, the fermionic sources.

\end{document}